\documentclass[aps,prl,twocolumn,amsmath,amssymb,showpacs]{revtex4}%
\usepackage{graphicx}
\usepackage{dcolumn}
\usepackage{bm}
\usepackage{color}
\usepackage{amsmath}
\usepackage{epstopdf}
\usepackage{fancybox}
\usepackage{amsfonts}
\usepackage{amssymb}%
\usepackage[toc,page]{appendix}
\providecommand{\U}[1]{\protect \rule{.1in}{.1in}}
\begin{document}
\title{Spin Superfluidity in Biaxial Antiferromagnetic Insulators}
\author{Alireza Qaiumzadeh}
\affiliation{Department of Physics, Norwegian University of Science and Technology, NO-7491 Trondheim, Norway}
\author{Hans Skarsv{\aa}g}
\affiliation{Department of Physics, Norwegian University of Science and Technology, NO-7491 Trondheim, Norway}
\author{ Cecilia Holmqvist}
\affiliation{Department of Physics, Norwegian University of Science and Technology, NO-7491 Trondheim, Norway}
\author{Arne Brataas}
\affiliation{Department of Physics, Norwegian University of Science and Technology, NO-7491 Trondheim, Norway}
\begin{abstract}
Antiferromagnets may exhibit spin superfluidity since the dipole
interaction is weak. We seek to establish that this phenomenon occurs
in insulators such as NiO, which is a good spin
conductor according to previous studies. We investigate non-local spin transport in a planar
antiferromagnetic insulator with a weak uniaxial anisotropy. The
anisotropy hinders spin superfluidity by creating a substantial
threshold that the current must overcome. Nevertheless, we show that applying a high magnetic field removes this obstacle near the spin-flop transition of the antiferromagnet. Importantly, the spin superfluidity can then persist across many micrometers, even in dirty samples.
\end{abstract}
% Antiferromagnetics, dynamics of Magnetization, magnetic properties of Films, spin transport effects
\date{\today}
\maketitle

%%%%%%%%%%%%%%%%%%%%%%%%%%%%
%%%% fig:set-up
%%%%%%%%%%%%%%%%%%%%%%%%%%%%
\begin{figure}[t]
\includegraphics[width=8.6cm]{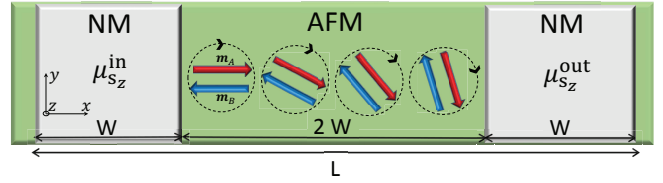}
\caption{SSF in a biaxial AFM insulator. The left and right normal
  metals (NMs) act as a spin injector and spin detector,
  respectively. The easy plane is the $xy$-plane and the easy axis is the $x$-direction. SSF occurs when the spins are tilted out of the easy plane and start to rotate around the hard axis with a spatially varying phase. }
\label{fig-setup}
\end{figure}
%%%%%%%%%%%%%%%%%%%%%%%%%%%%%

\textit{Introduction.---} Achieving long-range spin transport is
essential in spintronics. In metals, conduction electrons can carry
spin information. The spin-diffusion length is generally less than a
few hundred nanometers and often as short as a couple of
nanometers. However, in ferromagnets, there are additional transport
channels via spin excitations, typically in the form of spin waves. In
ferromagnetic insulators, the absence of noisy itinerant carriers
implies less dissipation such that magnons can traverse distances up
to several microns \cite{magnon-vanWees}. Magnetic low-damping
insulators in which new spin transport mechanisms can exist are of interest and can be promising candidates in low-dissipation spintronics.

Antiferromagnets (AFMs) have ordered spin configurations, but there is
no net magnetization at equilibrium. New observations and advances in
our understanding have stimulated increased interest in AFM
spintronics \cite{Ivanov,AFM-Jungwirth,AFM-Manchon,Sekine}. AFMs produce no
stray fields that can influence other elements. There are more known
high-temperature AFM insulators and semiconductors than their ferromagnetic counterparts.
AFMs exhibit transport properties similar to those of
ferromagnets. Some of these features are anisotropic magnetoresistance
\cite{AFM-AMR}, giant magnetoresistance \cite{AFM-GMR}, the large
anomalous Hall effect in non-collinear AFMs \cite{AFM-AHE}, and the spin Hall effect (SHE) \cite{AFM-SHE}. There are also recent investigations of the spin Seebeck effect in AFMs \cite{AFM-SSE-exp0, AFM-SSE-exp1, AFM-SSE-exp2, AFM-SSE-exp3}. Additionally, there are observations of spin transport in AFMs via spin pumping from an adjacent ferromagnet into AFMs \cite{AFM-SP-exp1, AFM-SP-exp2, AFM-SP-exp3, AFM-SP-theory}. In these experiments, it is possible that (evanescent) magnons carry the spin current \cite{Ivanov-AFM}.  A unique aspect of AFMs is that it is possible to trigger ultra-fast THz dynamics of the AFM order parameter via charge \cite{AFM-charge1, AFM-charge2} and spin currents \cite{Ran-Cheng}, magnons \cite{AFM-magnon}, spin-orbit torques \cite{AFM-SOT}, light \cite{AFM-light}, and spin Hall oscillators \cite{AFM-SHNO}.  Furthermore, magnon Bose-Einstein condensation (BEC) occurs in AFM insulators \cite{AFM-BEC, Eirik, Pokrovsky-YIG}.

In this Letter, we investigate spin transport via spin superfluidity (SSF)
in AFM insulators.
%Experiments have demonstrated that NiO is a good spin conductor \cite{AFM-SP-exp1, AFM-SP-exp3}.
We focus on NiO as a prototypical biaxial AFM insulator \cite{NiO-parameters1,
  NiO-parameters2, NiO-parameters3}. Crucially, the additional
anisotropy normally hinders SSF \cite{Sonin, Arne-book,
  MacDonald-book}. However, we demonstrate that near the spin-flop
transition \cite{spin-flop} superfluid behavior
still emerges and long-range spin transport beyond micrometers is feasible. This makes NiO and other biaxial AFMs promising for the first explicit experimental demonstration of SSF in magnetic materials.

Superfluidity is a dissipationless flow mediated by soft
Nambu-Goldstone boson modes \cite{Volovik}. Models of superfluidity
typically use a complex scalar field with global U(1) symmetry.  The
superfluid velocity is proportional to the gradient of the condensate
phase \cite{Volovik}. Halperin and Hohenberg demonstrated an analogy
between the spin dynamics in planar magnetic systems and the
hydrodynamic behavior of ideal liquids \cite{Halperin}. In a series of
seminal works, Sonin extended the concept of superfluidity of
electron-hole pairs \cite{e-h-SF, e-h-RMP} to spin systems and
introduced the notion of SSF \cite{Sonin}. In this scenario, SSF involves a $2\pi$-rotation of spins in a planar magnet.

However, some of us have recently demonstrated that in planar
ferromagnets, even as thin as 5 nm, the long-range dipole
interaction destroys SSF based on the proposed mechanism
\cite{Hans-FMSF}. Superfluidity reappears in synthetic AFM systems
\cite{Hans-FMSF}. Since the dipole interaction is negligible in AFMs,
we further explore SSF in AFMs \cite{Sonin,Ivanov-SSF-AFM,Yaroslav-SSF-AFM}.

The azimuthal angle of the order parameter, $\phi$, and the out-of-plane component of the total magnetization, $m_z$, can describe SSF \cite{Halperin,Sonin,Ivanov-SSF-AFM,Yaroslav-SSF-AFM,Hans-FMSF,Pokrovsky-YIG}. They are conjugate variables for AFMs. In the superfluid phase, $m_z$ is the superfluid density and $\phi$ is the phase of the condensate. The transverse component of the order parameter precesses. The spatial gradient of the superfluid phase is proportional to the spin supercurrent.

\textit{Setup.---} We consider a quasi-one-dimensional (1D) biaxial
AFM insulator. There is a strong hard-axis anisotropy and a weaker
easy-axis anisotropy. Two metallic layers attach at the left and the
right of the antiferromagnetic sample, as shown in
Fig.~(\ref{fig-setup}). Inducing SSF requires the spin accumulation to
be polarized along the hard anisotropy axis. The spin-valve structure
proposed in Ref.\ \cite{Hans-FMSF} can meet this requirement.  It is also possible to use a heavy metal to create a spin current via SHE. In the latter case, there must be a finite angle between the hard axis and the interface normal to ensure a significant superfluid spin density.

The width of both leads is $W$, and the separation between them is $2 W$. The total
system length is $L=4W$.  A spin accumulation generated by one of the
two mentioned injection methods induces a spin current from the
left. In turn, the spin current exerts a torque on the spins in the
AFM and causes precession. A small spatial gradient of the phase of
the precession governs superfluidity. Finally, spin pumping into the
right lead causes a spin accumulation therein. The resulting spin
accumulation can be measured using either a spin-valve structure or the inverse SHE.

\textit{Spin dynamics and stability criteria.---}
Superfluidity can be described semiclassically. We assume an AFM with two equivalent magnetic sublattices. The unit vectors along the directions of the magnetic moments are $\bm{m}_A(\bm{r},t)$ and $\bm{m}_B(\bm{r},t)$. At equilibrium, $\bm{m}_A(\bm{r},t)$ and $\bm{m}_B(\bm{r},t)$ are antiparallel. We introduce the magnetization, $\bm{m}=(\bm{m}_A+\bm{m}_B)/2$, and the staggered order parameter, $\bm{n}=(\bm{m}_A-\bm{m}_B)/2$. The effective total free energy density, see Eq. (S2) \cite{Suppl-Mat}, is given by \cite{Ivanov, units}
\small
\begin{equation}
f=\lambda^2\omega_{\mathrm{\|}}(\bm{\nabla} \bm{n})^2+\omega_{\bot}(\bm{n} \cdot \hat{z})^2-\omega_{\|}(\bm{n} \cdot \hat{x})^2+\frac{\omega^2_\mathrm{H}}{2\omega_{\mathrm{ex}}} (\bm{H} \cdot \bm{n})^2, \,
\label{free-energy-n}
\end{equation}
\normalsize
where $\lambda=\sqrt{\mathcal{A}/\omega_{\|}}$ is the domain wall (DW) length, $\mathcal{A}$ is the exchange stiffness, $\omega_{\|}$ is the uniaxial anisotropy, $\omega_{\bot}$ is the hard-axis anisotropy, $\omega_{\mathrm{ex}}$ is the homogeneous exchange energy, and $\omega_\mathrm{H}$ the Zeeman energy induced by an external magnetic field in the $\bm{H}$ direction. The exchange stiffness, $\mathcal{A}$, and the homogeneous exchange energy, $\omega_\mathrm{ex}$, are related via $\omega_{\mathrm{ex}}/\mathcal{A}=2 D/d^2$, where $D$ and $d$ are the spatial dimension and the lattice constant, respectively. The magnetization is a slave variable of the staggered order parameter, $\bm{m}=\dot{\bm{n}}\times\bm{n}/4\omega_{\mathrm{ex}}+\omega_\mathrm{H}\bm{n}\times (\bm{H}\times \bm{n})/2 \omega_{\mathrm{ex}}$ \cite{Ivanov, AFM-charge1,AFM-charge2}.

We consider an external magnetic field along the easy axis.  Eq.\ \eqref{free-energy-n} highlights that a critical magnetic field $H_c=\omega^c_{\mathrm{H}}/\gamma=\sqrt{2\omega_{\mathrm{ex}}\omega_{\|}}/\gamma$, compensates the uniaxial anisotropy. This peculiarity of AFM systems is known as the spin-flop transition \cite{spin-flop}. At the spin-flop transition, in this model, the free energies of biaxial AFMs become similar to those of planar AFMs with U(1) symmetry. However, the focus on this feature remains insufficient to prove, or fully understand, the range of validity of SSF.

The additional feature is that the magnetic field continues to influence the dynamic part of the AFM Lagrangian. This influence is via a gyrotropic term that breaks the Lorentz-invariant properties of AFMs,  $\mathcal{L}_{\mathrm{kin}}[\bm{n}]=\big(\dot{\bm{n}}^2+4\omega_{\mathrm{H}}\dot{\bm{n}}\cdot \bm{n} \times \bm{H} \big)/(8\omega_{\mathrm{ex}})$. It is the total Lagrangian density, $\mathcal{L}=\mathcal{L}_{\mathrm{kin}}-f$, that determines the dynamics of the N\'{e}el vector. In the presence of the spin transfer torque exerted by the spin accumulation in the left lead $\bm{\mu}_s^\mathrm{in}$ and the dissipation, see Eq. (S1) \cite{Suppl-Mat}, the dynamics of the staggered field is
\small
\begin{align}
&\bm{n}\times\big[\ddot{\bm{n}}-8\lambda^2\omega_{\mathrm{ex}}\omega_{\|} \nabla^2 \bm{n}-4 \omega_\mathrm{H} \bm{H}\times \dot{\bm{n}}+8\omega_{\mathrm{ex}}\omega_{\bot}n_z\hat{z}\nonumber\\&-8\omega_{\mathrm{ex}}\bar{\omega}_{\|}n_x\hat{x}+8 \alpha\omega_{\mathrm{ex}} \dot{\bm{n}}-4 \alpha_{\mathrm{SP}}\omega_{\mathrm{ex}}\bm{n} \times \bm{\mu}_s^\mathrm{in}\big]=0 \,
\label{EQ-n}
\end{align}
\normalsize
where $\bar{\omega}_{\|}=\omega_{\|}-\omega^2_{\mathrm{H}}/(2\omega_{\mathrm{ex}})$ is the effective easy-axis anisotropy energy that has been renormalized by the magnetic field, and the damping $\alpha(\bm{r})=\alpha_\mathrm{G}+\alpha_{\mathrm{SP}}(\bm{r})$, is sum of the Gilbert damping $\alpha_\mathrm{G}$ and the local damping enhancement $\alpha_{\mathrm{SP}}(\bm{r})$.

To study spin transport in the setup of Fig.~\eqref{fig-setup}, we
consider 1D solutions of the linearized equation of
motion of Eq. \eqref{EQ-n}. We use spherical coordinates for the staggered order parameter field $\bm{n}=\{\sqrt{1-n_z^2} \cos\phi, \sqrt{1-n_z^2} \sin\phi, n_z\}$, where $n_z$ is the out-of-plane deviation.

First, we consider the static regime to find the critical current required to trigger SSF and the Landau criteria for the breakdown of SSF. To the linear order in $n_z$, we find
\small
\begin{subequations}
\begin{align}
&\bar{\omega}_\| \sin 2 \phi-2\lambda^2\omega_{\|} \partial_x^2 \phi=-\alpha_{\mathrm{SP}}\mu_z,\\
&({\omega}_\bot+\bar{\omega}_\| \cos^2\phi)n_z-\lambda^2 {\omega}_\| \partial_x^2 n_z=0 \, .
\end{align}\label{EQ-theta-phi}
\end{subequations}
\normalsize
where the driving force is  $\mu_z=\mu_{s_z}^{\mathrm{in}}\Theta(W-x)$
and $\Theta$ is the Heaviside function. One solution of
Eq.\  (\ref{EQ-theta-phi}) is a homogeneous state. The spins are then
in the easy plane, $n_z=0$, and the azimuthal angle is governed by the
STT, $\sin
2\phi_0=-(V_L/V_0)\alpha_{\mathrm{SP}}\mu_{s_z}^{\mathrm{in}}/\bar{\omega}_{\|}$,
where $V_L$ is the partial AFM volume below the left lead.  The static
macrospin solution becomes unstable when the spin-transfer torque is
sufficiently large, $|\alpha_{\mathrm{SP}} \mu_{s_z}^{\mathrm{in}}|> V_0
\bar{\omega}_{\|}/V_L$. Consequently, in the
presence of a finite effective uniaxial anisotropy
$\bar{\omega}_{\|}$, triggering SSF requires a large spin accumulation
when the spin-pumping-enhanced damping $\alpha_{\mathrm{SP}}$ is
small. It is therefore essential to reduce the effective uniaxial anisotropy by an external magnetic field.
In this regime, there are also two types of spatially varying solutions of Eq.\ (\ref{EQ-theta-phi}). An in-plane homogeneous spiral state is a stable state when  $\bar{\omega}_{\|}=0$ while a kink-like state becomes more stable in $\bar{\omega}_{\|}\neq0$ \cite{MacDonald-book,Bloch-Domain}.
%Allowing for a spatial variation of the staggered field, but in the absence of the both effective easy axis anisotropy and the magnetic field, Eq. \eqref{EQ-theta-phi} have two kinds of solutions. One solution is a spiral state $n_z=0$ and $\phi(x)=\phi_0+q x$, where $q$ the spiral wave vector.
%Other static solution of this equation is a Bloch DW solution, in which the spins in the center of the wall are out of plane. This Bloch wall solution is not the ground state of the system but might be excited thermally as already discussed in \cite{Yaroslav-phase-slip1} in ferromagnets. This causes a phase slip that destroys the spin supercurrent. The easy axis anisotropy prevents thermal phase slips since the Bloch DWs in 1D and vortices in 2D are no longer a solution of Eq.\ (\ref{EQ-theta-phi}).

Next, we allow the spins to vary in time. The dynamics of the conjugate variables, up to the linear order in $n_z$ and the derivatives of $\phi$, are described by
\small
\begin{subequations}
\begin{align}
\dot{\phi}=&-4 \omega_{\mathrm{ex}} m_z- 2 \omega_\mathrm{H} n_z \cos\phi, \label{EQ-phidot}\\
\dot{m}_z=&-2\lambda^2\omega_{\|} \partial_x^2 \phi+\bar{\omega}_{\|} \sin2\phi+2\alpha_\mathrm{G} \dot{\phi}\nonumber\\&+\frac{\omega_\mathrm{H}}{2\omega_{\mathrm{ex}}}(\dot{n}_z \cos\phi+n_z\dot{\phi}\sin\phi). \label{EQ-mzdot}
\end{align}
\end{subequations}
\normalsize
Eq. (\ref{EQ-mzdot}) is a continuity equation for the out-of-plane component of the magnetization. In an ideal planar AFM, $\omega_{\|}=\omega_\mathrm{H}=\alpha_\mathrm{G} =0$; thus, the $z$-component of the magnetization is conserved and Eqs. (\ref{EQ-phidot}) and (\ref{EQ-mzdot}) are similar to the Josephson supercurrent equations \cite{Sonin, MacDonald-book}.

Finally, the dynamics of $\phi$ and $n_z$ read as
\small
\begin{subequations}
\begin{align}
&\ddot{\phi}-v_c^2(\partial_x^2 \phi -\frac{1}{2\bar{\lambda}^2}\sin 2\phi)+8 \alpha_\mathrm{G}\omega_{\mathrm{ex}} \dot{\phi}+4\omega_\mathrm{H}\dot{n}_z\cos\phi=0, \label{EQ-phi}\\
&\ddot{n}_z-v_c^2 \partial_x^2 n_z+8 \alpha_\mathrm{G}\omega_{\mathrm{ex}} \dot{n}_z+8 \omega_{\mathrm{ex}}n_z (\omega_\bot+\bar{\omega}_{\|}\cos^2\phi)\nonumber\\&-4 \omega_\mathrm{H}\cos\phi\dot{\phi}=0, \label{EQ-nz}
\end{align}
\end{subequations}
\normalsize
where $\bar{\lambda}=\sqrt{\mathcal{A}/\bar{\omega}_{\|}}$ is the effective DW length and $v_c=2\lambda\sqrt{2\omega_{\mathrm{ex}}\omega_{\|}}$ is the effective velocity of "light".  The AFM spin wave velocity, $v_c$, is significantly larger than its ferromagnetic counterpart by a factor of $\sqrt{\omega_{\mathrm{ex}}/\omega_\bot}$ \cite{MacDonald-book}.

%Only after developing these equations, Eqs. (\ref{EQ-phidot})-(\ref{EQ-nz}), can we explain why we can disregard the remaining magnetic field influence of the gyrotropic term.
The gyrotropic term in the Lagrangian causes the last terms in the dynamical equations, Eqs. (\ref{EQ-phidot})-(\ref{EQ-nz}). The gyroscopic term couples the dynamics of the condensate phase to the out-of-plane component of the order parameter. However, this term is proportional to $\dot{n}_z$, \ref{EQ-phi}, and might be disregarded when a strong hard-axis anisotropy suppresses the out-of-plane dynamics \cite{Suppl-Mat}.

%%%%%%%%%%%%%%%%%%%%%%%%%%%%
%%%% fig:J-H
%%%%%%%%%%%%%%%%%%%%%%%%%%%%
\begin{figure}[t]
\includegraphics[width=8cm]{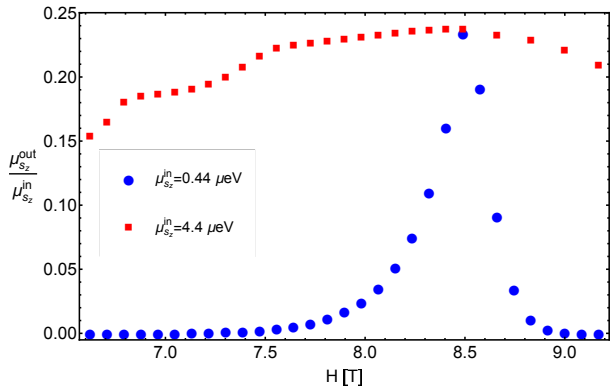}
\caption{The ratio between the output spin accumulation and the input spin accumulation as a function of the applied magnetic field for two different input spin accumulations.}
\label{J-H}
\end{figure}
%%%%%%%%%%%%%%%%%%%%%%%%%%%%%

Eqs. (\ref{EQ-phi}) and (\ref{EQ-nz}) decouple if
$\omega_\mathrm{H}={\omega}_{\|}=0$. Then, Eq. (\ref{EQ-phi})
determines gapless phase excitations, $\omega_\phi=v_c k$. Conversely,
Eq. (\ref{EQ-nz}) implies that excitations of ${n}_z$ have a gap,
$\omega_{n_z}=\sqrt{v_c^2 k^2+8 \omega_\bot \omega_{\mathrm{ex}}}$. Since the gap is large, it is considerably
more difficult to excite dynamic variations in $n_z$. This is different in planar ferromagnets, where there is only one gapless magnon mode with linear dispersion
\cite{Yaroslav-FMSF,Ivanov-SSF-AFM,Yaroslav-SSF-AFM}. In the
considered regime, $\omega_\mathrm{H}={\omega}_{\|}=0$, the steady-state solution is $\phi(x,t)=\phi(x)+\Omega t$. The precession
frequency is determined by the driving force of
Eq.\ (\ref{EQ-theta-phi}) such that
$\alpha_\mathrm{G}\Omega=\lambda^2\omega_{\|}\partial^2_x\phi$. The
$z$-component of the magnetization is conserved. The continuity
equation in the low Gilbert damping limit reads as $M_s
\dot{m}_z=-\partial_x J_{s_z}$, where $M_s$ is the saturation
magnetization and the spin supercurrent density is $J_{s_z}=2 M_s
\lambda^2\omega_{\|}\partial_x\phi(x)$ \cite{Sonin,Ivanov-SSF-AFM,
  Yaroslav-SSF-AFM}.
%As we have already discussed, this dependence of the
%spin superfluid current on the gradient of the condensate phase is similar
%to the characteristic of the superfluid hydrodynamics of conventional superfluids.

In the presence of a finite uniaxial anisotropy and magnetic field, the superfluid density $m_z$ is no longer conserved. Nevertheless in the weak dissipation limit and at the spin-flop transition, the angular momentum is conserved, see Fig. S1 \cite{Suppl-Mat}.

%Note that the last term in the continuity equation Eq. (\ref{EQ-mzdot}), which originated from the gyrotropic term in the Lagrangian, is tiny and can be neglected within the exchange approximation. This term leads to a very small nutation in the $2\pi$-rotation of the order parameter, see Fig. S1 \cite{Suppl-Mat}.

When $\bar{\omega}_\parallel\neq 0$, Eq. (\ref{EQ-phi}) has a kink-like soliton solution for the condensate phase, $\phi(x,t)=2 \tan^{-1}\big[\exp\big(\pm(x-x_0-v t)/(\bar{\lambda}\sqrt{1-(v/v_c)^2})\big)\big]$, where $x_0$ is the arbitrary DW center position and $v$ is the DW velocity. An inhomogeneous state then becomes stable. The DW velocity is determined by the driving force, resulting in $v\sim \pm W \alpha_{\mathrm{SP}} \mu_s^{\mathrm{in}}/(2 \pi\alpha_\mathrm{G}) < v_c$. The traveling solitons generate an AC signal on top of a DC output in the detector. At the spin-flop transition, $\bar{\lambda}^{-1}=0$, only the DC component of the signal survives and the system exhibits perfect SSF.

Let us investigate the conditions for SSF stability. We will find two
criteria that also exist in similar forms in ferromagnetic SSF
\cite{Sonin}. To this end, we consider the limit of a large hard-axis
anisotropy; then, $n_z$ is very small, and the total
free energy becomes $f \sim \lambda^2 \omega_\parallel
(\partial_x\phi)^2
+[\omega_\perp-\lambda^2\omega_\parallel(\partial_x\phi)^2+\bar{\omega}_\parallel\cos^2\phi]n_z^2-\bar{\omega}_\parallel\cos^2\phi$. At
the spin-flop transition, $\bar{\omega}_\parallel=0$, the free energy
implies that the SSF remains stable provided that $\partial_x \phi<\sqrt{\omega_\perp/(\lambda^2 \omega_\parallel)}$. This upper critical limit for the gradient of the condensate phase is analogous to the Landau criterion in conventional superfluidity. A finite effective anisotropy increases the upper critical limit. Another necessary condition is that the spin supercurrent must be spatially uniform. The steady-state solution of Eq. (\ref{EQ-phi}) gives rise to an approximately uniform spin supercurrent only when $\partial_x \phi\gg 1/\bar{\lambda}$ \cite{Sonin}.

\textit{Numerical results and discussion.---}
To establish SSF, we numerically solve the coupled Landau-Lifshitz-Gilbert (LLG) equations for sublattice magnetization $\bm{m}_A$ and $\bm{m}_B$, Eq. (S1) \cite{Suppl-Mat}, for the setup in Fig. (\ref{fig-setup}). In the numerical calculations, we use parameters for the prototypical AFM insulator NiO \cite{NiO-parameters0,NiO-parameters1, NiO-parameters2, NiO-parameters3}. The spin-flop transition occurs at a critical external magnetic field of $H_c \sim 8.5$ T \cite{spinflop-NiO}. It was experimentally shown that at near room temperature $H_c$ is reduced to approximately 2 T in NiO \cite{spinflop-NiO}. We also estimate, $\omega^c_{\mathrm{H}}/2\omega_{\mathrm{ex}} \sim 10^{-3}$, and thus, the $z$-component of the spin current is approximately conserved at the spin-flop transition Eq. (\ref{EQ-mzdot}).
%%%%%%%%%%%%%%%%%%%%%%%%%%%%
%%%% fig:Jin-Jout
%%%%%%%%%%%%%%%%%%%%%%%%%%%%
\begin{figure}[t]
\includegraphics[width=8cm]{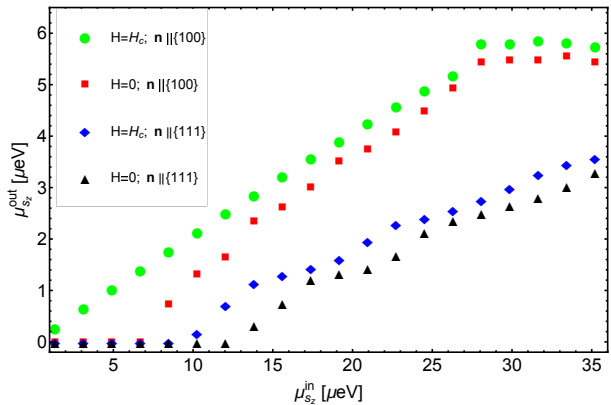}
\caption{The output spin accumulation (in the right lead) as a function of the input spin injection (in the left lead).}
\label{J-alpha}
\end{figure}
%%%%%%%%%%%%%%%%%%%%%%%%%%%%%

The spin accumulation pumped into the right lead, in the limit of low
spin-memory loss, is $\bm{\mu}_s^{\mathrm{out}}=-(1 /2) \sum_i
\bm{m}_i \times \dot{\bm{m}}_i$
\cite{Arne}. When the hard-axis anisotropy is large, the $z$-component
of the pumped spin accumulation is given by $\mu_{s_z}^{\mathrm{out}}
\approx -\dot{\phi}$. In Fig. (\ref{J-H}), we plot the normalized spin
accumulation in the right lead versus the input spin accumulation in
the left lead, $\mu_{s_z}^{\mathrm{out}}/\mu_{s_z}^{\mathrm{in}}$, as
a function of the applied magnetic field along the $x$-direction
$H$. The system length is $L=0.75 \mu$m.  We consider two STT
amplitudes: $\mu_{s_z}^{\mathrm{in}}=0.44 \mu$eV and
$\mu_{s_z}^{\mathrm{in}}=4.4 \mu$eV. In both cases, the maximum
spin-transport efficiency occurs at the spin-flop transition. When the
STT energy is smaller than the easy-axis energy, the output signal
vanishes rapidly when the applied magnetic field deviates from the
critical field. Naturally, we still find large signals when the input
STT energy is larger than the uniaxial energy. In the latter case, the
STT is strong enough to overcome the pinning arising from the
effective uniaxial anisotropy and triggers coherent spin dynamics even
at lower external magnetic fields. This field-dependent behavior can
be used to experimentally distinguish the spin current arising from usual magnons and the spin current carried by SSF.

In Fig. (\ref{J-alpha}), we plot the detected spin accumulation as a
function of the STT amplitude for a system size $L=0.75\mu$m. We show
the results with and without an applied critical magnetic field. When
there is no magnetic field, the spins are pinned by the uniaxial
anisotropy. In this case, the STT amplitude should be larger than a
threshold that is proportional to the effective uniaxial anisotropy
energy $\bar{\omega}_x$ to induce spin transfer. Near the spin-flop transition, this limit vanishes due to the restoration of the U(1) symmetry. At the spin-flop transition,  higher order anisotropy terms may be present in the free energy; see for example, Ref. \cite{Ivanov}. In this case, there is no longer a perfect U(1) symmetry. This residual anisotropy can easily be overcome by a small spin accumulation. Above the threshold, the ratio between the pumped and injected spin currents is linear in the input spin accumulation but only up to another critical value determined by the Landau criterion. There is no longer a typical superfluidity behavior beyond this point.

%%%%%%%%%%%%%%%%%%%%%%%%%%%%%
%%%%% fig:J-time
%%%%%%%%%%%%%%%%%%%%%%%%%%%%%
\begin{figure}[t]
\includegraphics[width=8cm]{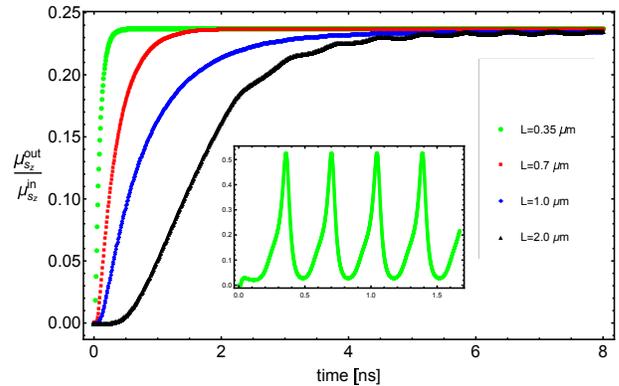}
\caption{The time evolution of the output spin accumulation for
  different length scales at the spin-flop transition $H=H_c$ when
  $\mu^{\mathrm{in}}_{s_z}=2 \mu$eV. Inset: Below the spin-flop
  transition $H\simeq 6.6$ T when $\mu^{\mathrm{in}}_{s_z}=4.4 \mu$eV.}
\label{J-L}
\end{figure}
%%%%%%%%%%%%%%%%%%%%%%%%%%%%%%

The magnetic moments in NiO are ferromagnetically aligned in the
$\{111\}$ planes. The adjacent planes are antiferromagnetically
coupled. Thus, the easy planes are not parallel to, e.g., a surface
along $\{100\}$. We therefore also explore the likely effect of a
finite angle between the easy plane of the NiO layer and a Pt/NiO
interface. To this end, we rotate the easy plane in our numerical
calculations while maintaining the applied magnetic field in the
$x$-direction. Fig. (\ref{J-alpha}) demonstrates that SSF remains
feasible at finite angles. Only the component of the magnetic field parallel to the uniaxial anisotropy reduces the effective anisotropy. Then, in the presence of a finite angle between the external magnetic field and the uniaxial anisotropy $\theta$, the critical magnetic field is increased to $H_c/\cos \theta$.

We plot the time evolution of the spin current for different length
scales in Fig. (\ref{J-L}). At the spin-flop transition point, even in
a dirty sample where $\alpha_\mathrm{G}\sim 6.8\times 10^{-3}$
\cite{NiO-parameters1}, SSF persists up to a few micrometers.
%In clean samples, where $\alpha_\mathrm{G}\sim 2\times10^{-4}$
%\cite{NiO-parameters3}, the range of SSF can be further increased
%dramatically.
This micron size range of the spin transport by SSF is
considerably larger than the damping decay length of magnons in NiO,
$\lambda_{\mathrm{G}}=
v_c/(4\alpha_{\mathrm{G}}\omega_{\mathrm{ex}})\sim 40$ nm.
Fig. (\ref{J-L}) also shows that when the system size increases, the
transient time increases. In general, in the presence of a uniaxial anisotropy and a strong STT amplitude, the output spin accumulation has both AC components in the GHz regime together with a DC component. As discussed, the AC signal is a consequence of injecting kink-like solitons from the left lead. There is a reduction of the AC signal near the spin-flop transition. The signal is purely DC exactly at the transition point.

Experimentally, magnons and SSF contribute to the
output spin accumulation. Their contributions can be distinguished
either by changing the strength and direction of the magnetic field or
by changing the sample size. Magnons decay exponentially with the
system size, whereas we expect a very small algebraic decay for SSF in our setup.

\textit{Conclusion.---}
NiO is a biaxial AFM without a U(1) symmetry. This
material can also have a significant Gilbert damping. These features
appear to be detrimental to long-range spin transport via both magnons and SSF. Nevertheless, we demonstrate that NiO and other biaxial AFMs are good candidates for observing SSF over micrometer length scales. SSF behavior is dramatically improved around the spin-flop transition, which can be reached by applying an external magnetic field. SSF can be observed in standard non-local spin-transport setups and reach distances beyond micrometers.

The research leading to these results has received funding from the European Research Council via Advanced Grant number 669442 ``Insulatronics''. We would like to thank M. Kl\"{a}ui and J. Cramer for useful discussions.


\begin{thebibliography}{9}
\bibitem {units}
Throughout the paper we use the natural units $\hbar=c=1$, where $\hbar$ is the Planck constant and $c$ is the light velocity in the vacuum.
\bibitem {Bloch-Domain}
In the static limit of Eq. (\ref{EQ-n}), there is also a Bloch domain wall-like state solution when $\bar{\omega}_\parallel=0$. This Bloch state can be excited thermally as discussed for ferromagnets \cite{Yaroslav-phase-slip1}. Thermal excitations then cause a phase slip that destroys the spin supercurrent.
\end{thebibliography}
\end{document}